\pdfoutput=1

\documentclass[12pt]{article}

\usepackage{hyperref}
\hypersetup{
  letterpaper,
  colorlinks,
  urlcolor=blue,
  citecolor=blue,
  linkcolor=red,
  pdfpagemode=none,
  pdftitle={Curriculum Vitae},
  pdfauthor={Subhadeep Mukhopadhyay},
  pdfsubject={Curriculum Vitae},
}% to color the links

\usepackage[intlimits]{amsmath} % necessary for \tag [] for limits in the inegral proper place
\usepackage{amsthm,mathrsfs}    % same
\usepackage[comma,authoryear]{natbib} 
\numberwithin{equation}{section}

\numberwithin{equation}{section}
\usepackage{amssymb,amsmath}
\usepackage{subfigure}
\usepackage{bm}
\usepackage{graphicx,graphics,epsfig}
\usepackage{url}
\usepackage{multirow}

\theoremstyle{plain} 
\newtheorem{thm}{Theorem}
\newtheorem{coro}{Corollary}
\newtheorem{prop}{Proposition}

\theoremstyle{definition} 
\newtheorem{example}{Example}

\theoremstyle{remark} %italic style
\newtheorem{remark}{Remark} % \begin{remark}  \end{remark}

\newcommand{\bthm}{\begin{thm}}
\newcommand{\ethm}{\end{thm}}

\newcommand{\bpf}{\begin{proof}}
\newcommand{\epf}{\end{proof}}

\usepackage[compact,small]{titlesec}
\usepackage{deep-macro}

\begin{document}
 
\begin{center}
{\Large {\bf Quantile Based Variable Mining : \\ Detection, FDR based Extraction and Interpretation  }  }
\\[.4in]
Preprint version, 14 Dec, 2011 \\
 S. Mukhopadhyay, Emanuel Parzen and S.N.Lahiri\\
Department of Statistics, Texas A \& M University\\
College Station, TX  77843-3143\\[.4in]

{\bf ABSTRACT}\\
\end{center}
This paper outlines a unified framework for high dimensional variable selection for classification problems.
Traditional approaches to finding interesting variables mostly utilize only partial information through moments (like mean difference). On the contrary, in this paper we address the question of variable selection in full generality from a distributional point of view. If a variable is not important for classification, then it will have similar \textit{distributional} aspect under different classes. This simple and straightforward observation motivates us to quantify \textit{`How and Why'} the distribution of a variable changes over classes through CR-statistic. The second contribution of our paper is to develop and investigate the FDR based thresholding technology from a completely new point of view for adaptive thresholding, which leads to a elegant algorithm called CDfdr.
This paper attempts to show how all of these problems of detection, extraction and interpretation for interesting variables can be treated in a unified way under one broad general theme - \textit{comparison analysis}. It is proposed that a key to accomplishing this unification is to think in terms of the quantile function and the comparison density. We illustrate and demonstrate the power of our methodology using three real data sets.

%  Key aspects of our method are (i) \textit{Invariance}: No need to do perform normalization for microarray data, as our method is invariant under monotone transformation. (ii) \textit{Data Integration}: We can readily integrate data from different microarray platforms (DNA microarray technology/counting based technology like Massively parallel signature sequencing (MPSS)) as our transformation make the distribution of each gene nearly uniform. (iii) \textit{Quadratic Detector} : Our method generalizes linear rank statistics to accommodate possible higher order information;  (iv) \textit{Interpretation}: Through the shape of the comparison density. (v) \textit{Signal Extraction}: We have developed a novel FDR based threshold selection method and motivated it from goodness of fit perspective.
% This paper attempts to show how all of these problems can be treated in a unified way under one broad general theme \textit{comparison analysis}. It is proposed that a key to accomplishing this unification is to think in terms of quantile function and comparison density. It is shown that our method outperforms existing methods for variable selection.
%can be integrated 
\noindent ----------------------------------------------------------------\\
{\it Keywords and phrases:} Comparison density, Mid-distribution function, Variable selection, 
Ranking, Orthogonal Series Density Estimation, Score function, Wilcoxon Statistics, FDR, Pre-flattened smoothing  .
\newpage
\setcounter{equation}{0}
\section{Introduction}
 Consider a classification problem where we have a large number of predictor variables $X_1,X_2,\ldots,X_p$ and where $Y$ denotes the class labeling. Our goal is to find the most `influential' variables, and more importantly, to gain a deeper insight on the questions : \textit{why} and \textit{when} as well as \textit{how} a variable could be influential, through visualization. This extra piece of information can help applied researchers not only to identify the \textit{needle} in a haystack but to classify them according to their \textit{ impact types} and thus, provide specific hypotheses for further investigation. 
%Yet surprisingly all available methods are silent on this point. 

% \subsection{Data introduction} As a means of motivating our approach we will introduce two examples. Here is the brief introduction to the examples.
% \begin{example}(Hepatitis Data, \cite{hepa}) 
% 
% $p=19$ features was selected from $n=155$ samples on two different target classes ($n_1= 32$ and $n_2= 123$). We have t\textit{hree different types} of attributes continuous (eg. variable \# 14: Bilirubin amount), discrete (eg. variable \# 18: Protime) and categorical (eg.variable \# 2: Gender ) and lots of \textit{missing values }specially for variable \# 18, which presumably is a very important variable.
% \end{example}
% 
% \begin{example}(Prostate Cancer Data, \cite{prostate})
% 
%  This data set gives the expression levels of $p=6033$ genes for $n_1=50$
% normal tissues and $n_2=52$ prostate cancer tissues. The distinctive features of this data set compared to the previous one is that, here the expression values are highly contaminated with \textit{noise} and we occasionally face \textit{spurious large values}.
% \end{example}

\begin{figure*}[tpbh] 
%\vspace*{-.1in}
\begin{center}
\includegraphics[width=8.5in]{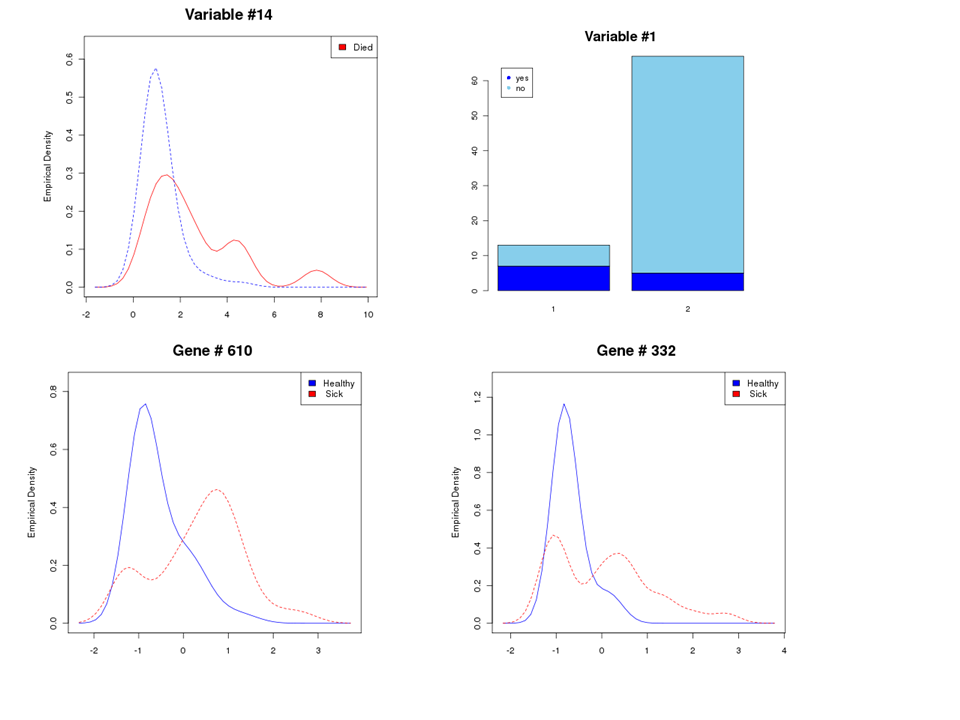}\end{center}
\vspace*{-.5in}
\caption{\emph{Important Variables are Different for Different reasons  : Panels A,B : give is the plots of the distributions of Variable \# 14 (continuous) and \# 1 (categorical) for hepatitis data set over two classes $Y=1$ (`Dead') and $Y=0$ (`Alive'). Panels C and D: show distribution of Genes \# 610 and \# 332 of the Prostate Cancer data, over the two classes $Y=1$ (`Cancer') and $Y=0$ (`Healthy'). It shows that the distinguishable discriminatory pattern is different for different variables .}} \label{pic:intro}
\vspace*{-.2in}
\end{figure*}

To capture the interesting variables, first we need to understand how a variable could be informative about the classes. For this we consider a few illustrative examples and attempt to define exactly what is meant by `interesting/uninteresting'.  
Figure \ref{pic:intro} gives an idea about discriminatory information (\textit{how and what}) a variable could render and set the stage for our method. We now highlight a few key aspect of Figure \ref{pic:intro} with regard to variable selection methodology 

\vspace{.5em}
\noi (a.)  \textit{Information: Distribution vs. Moments}.  Figure 1 in panel A, demonstrates the difference in location and scale of variable \# 14 (Bilirubin) of the Hepatitis data \citep{hepa} for the two classes, `dead' and `alive'. Note that the variability of Bilirubin is higher in the ``red'' class (`dead') compared to the blue one. Similarly, Panel C on the distribution of Gene \# 610 of the Prostate Cancer data (cf. Section 5) for the two classes (`sick' and `health') shows contrasting skewness and considerable departure from normality and Panel D on Gene \# 332 of the same data set clarifies the existence of contrasting \textit{tail behavior } and the presence of bi-modality, which gives a strong message about the presence of the disease ; This shows that each variable may exhibit a different kind of information and we want our methodology to be flexible enough to capture these effects. So, this automatically raises the issue as to how we can possibly identify the variables with different types of information and how we should measure those in a coherent manner. Thus we need a methodology that, lets the data determine the variables of importance to the greatest extend possible.

This paper aims to exploit the \textit{distributional} information of a variable over different classes by developing a set of score statistics to quantify that. This is in marked contrast to the current practice of identifying `interesting' of variable solely on the basis of a single \textit{moment} based information ( e.g., mean level change).

\begin{remark}[\texttt{Biological Significance}]
Not until very recently, biologist have noticed the importance of detecting the higher order information in genes. \cite{FI10} and \cite{Nature11}, have argued and confirmed the importance of detecting genes which are differentially expressed in terms of ``\textit{variability}'' as opposed to the traditional approach of discriminating on the basis of the ``\textit{mean'' change} for colon tumor. This is further confirmed by Figure \ref{pic:intro}, where we can easily see that \textit{diseased} class exhibit higher variability than the \textit{normal} class for all of the variables. Although most traditional variable selection  approaches use only partial discriminative information through the first order (i.e., through the mean level) to filter the important variables, our method is flexible enough to detect those genes which show increased variability and other specific distributional aspects (e.g., higher order moment information) in cancer tissues compared to healthy normal samples and thus, have the potential to find a new set of disease-related genes which was not \textit{previously anticipated}.
\end{remark}
 
\noi (b.) \textit{Nonparametric and Robust}. Our approach is fully nonparametric and robust. Further it can accommodate missing values (e.g., the Hepatitis data set that we refer to Fig. 1 contains lots of missing observations). Another useful quality of our methodology is that it is \textit{invariant under monotone transformations}, as a result of it works perfectly fine without any type of normalization techniques (specially for microarray data). The extended long tail of Gene \# 332 in panel C  and the omnipresence of noise in the microarray data necessitate robustness against departure from normality, which is guaranteed by our method.

\noi (c.) \textit{Unification and Integration}.  Our methodology provides a mechanism for handling continuous, discrete and categorical variables (Hepatitis data contains all of them) through an ingenious \textit{mid-distribution transformation}. This enables us to propose a unified method for finding  differentially expressed genes under different microarray platforms (DNA microarray technology/Counting based technology like Massively parallel signature sequencing (MPSS)/Next generation sequencing technology (RNASeq data)).  

\noi (d.) \textit{Characterizing the Information content}.   Each interesting variable has different information content for different reasons and that is why we seek a method that would shed light on \textit{how and why} a particular variable is important. One of the most important features of our method is its ability to offer \textit{additional interpretability and understanding} in scientifically meaningful terms through visualization.  This non-classical property of variable selection would help the scientist to get a fuller picture of what is happening.
\vspace{.2em}

\begin{figure*}[tpbh] 
\vspace*{-.1in}
\begin{center}
\includegraphics[width=6.2in]{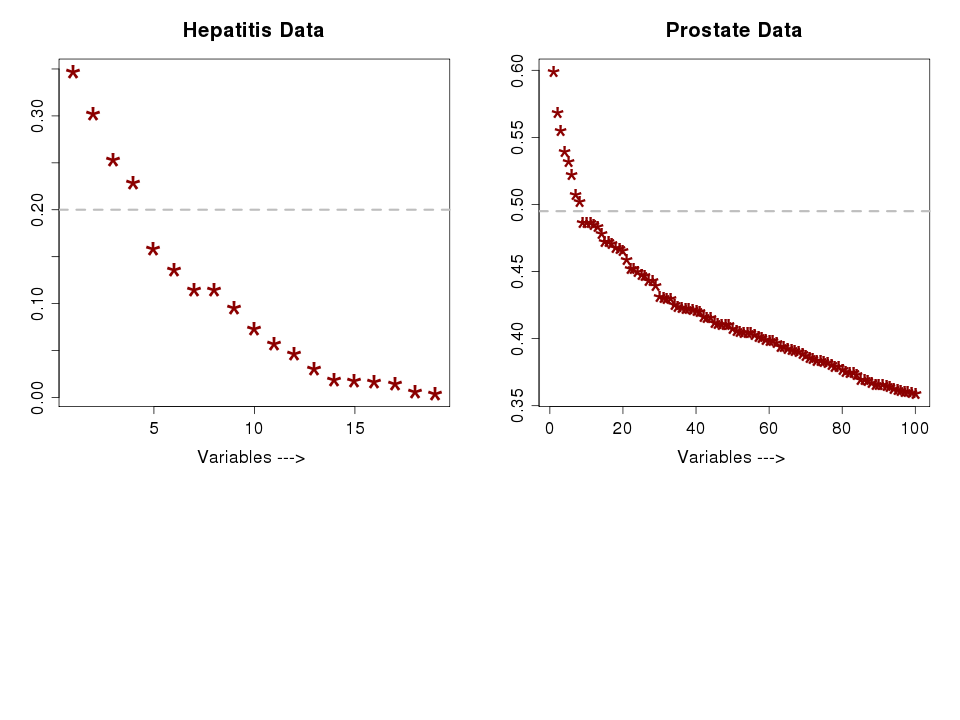}\end{center}
\vspace*{-1.85in}
\caption{\emph{Importance ordering  : Each panel plots the value of CR-statistics after sorting. Panel A deals with the Hepatitis study where we can see a natural gap, marked using a gray dotted line. Panel B: Prostate cancer study. We have only plotted top $100$ score statistics for clarity. Our CDfdr algorithm presented in the Section 6.2 detects such natural gaps or thus selects proper thresholds}} \label{pic:thres}
\end{figure*}

%  With this in mind, the first aim of this article to present a modern \textit{quantile based} methodology for variable selection which is (i) Nonparametric; (ii) Robust; (iii) Non-linear; (iv) works for all data types (v) accommodates missing observation. The most important feature of our method is it's ability to offer an \textit{additional interpretability and understanding} in scientifically meaningful terms through visualization.
In Sections 2-4 we discuss several key concepts and present a novel algorithm to compute and interpret such variable importance measures once we quantify the importance of each of the variables, we can rank the variables in a decreasing order as in Figure \ref{pic:thres}. Section 5-6, the main part of the paper, describes a simple and straight forward approach for FDR corrected threshold selection based on comparison density. It turns out that the threshold selection problem is intimately related to the problem of estimating the \textit{tail} of 
the \textit{ratio} of two density. In this respect, we have developed an elegant fully nonparametric approach to FDR based adaptive thresholding, which we shall refer to as the \textit{CDfdr Algorithm}. The key to accomplishing this also turns out to be expressing the FDR in terms of the comparison density, a fundamental concept introduced by \cite{parzen79}.

The main observation behind the construction of the proposed CDfdr based variable selection procedure is that \textit{directly} estimating the ratio of two density is more efficient and stable, compared to estimating them separately and then taking the ratio. We also employ an approach based on \textit{Pre-flattened smoothing of comparison density} which can efficiently capture the behavior of the tail of the ratio of two densities. This connection among the comparison density, the false discovery rate and the Pre-flattening approach is new. Our framework for threshold selection provides a new perspective that  could be considered as a general methodology for adaptive threshold selection for recovering signal from its noise corrupted version. 

\vspace{.7em}

In summary, we provide a fresh look at the problem of variable selection and address the issue using various key concepts from modern quantile based nonparametric methods.
In the first part (Section 2-4) of the paper, we describe the background concepts and attempt to build the detector based on CR-statistic which can potentially quantify the '\textit{informativeness}' of a variable. The second part (Section 5-6) is devoted to building the CDfdr algorithm to separate the '\textit{significant}' (signal) variables from the noise. We have made a special effort to illustrate our concepts through various examples. This is a very difficult task as we are to work with thousands of variables where each variable is special in its own right. In this article, we have limited ourselves to the independent case while remaining central to the topic at hand. We do this primarily to introduce the remarkably powerful foundational ideas, which have enormous potential to tackle high dimensional inference from a completely nonparametric and robust point of view.

\section{Background Concepts}
\subsection{Mid-Distribution Transform} \label{mrt}
\noi \textit{Definition and Properties.} 

\cite{parzen89} introduced the seminal concept of a mid-distribution function, which is calculated as a transformation of ranks of the tied data. Let $X$ be a random variable with distribution function $F(x) = \Pr(X \le x)$ and probability mass function (pmf) $p(x) = \Pr(X=x)$. The Mid-distribution function is defined as,
\beq 
\label{fmid} 
\Fm(x) =  F(x) - .5 p(x),  ~x\in \cR.
\eeq
When the sample consists of distinct values, $\Fm(x_j)= (R_j - .5)/n$, where $R_j$ is the rank of $x_j$ in the sample of size $n$. For the tied cases we use the average rank instead of $R_j$. Note that for continuous random variable X, we have $\Fm(X) = F(X) \sim U(0,1)$.

\noi \textit{A Small Example.} 

Table \ref{table:fm} shows the raw data along with the corresponding mid-rank transformed version for first $10$ sample values of variable \# 18 (Hepatitis data).

\begin{remark}[\texttt{Why Fmid-Transformation}]
(i) \textit{Unification}: The notion of mid-rank transformation enable us to deal simultaneously with discrete and continuous distribution. 
(ii) \textit{Robustness} : The rank based nature of the Fmid transformation makes it robust. (iii) \textit{Invariance Property \& Normalization}: The standard practice for normalizing microarray data is to take log-transformation (as a variance-stabilizing transformation) for normalization. But although log-transformation stabilizes the variance of high expression labels, it is known that this transformation increases the variance of the observations near the background. Our Fmid based \textit{nonlinear} transformation is \textit{invariant under monotone transformation} and we do not need to supply the calibrated (biased) microarray expression. Our method can directly work on the raw data and produce the same result. From another angle, Fmid based transformation could be considered as a universal normalization technique, regardless of the measurement error structure. 
\end{remark}

\noi \textit{How to Compute.}

We will start all our nonparametric statistical data analysis by taking mid-rank based \textit{nonlinear transformation} of the raw data $\xb$ .
Transforming the raw data $\xb = (x_1,x_2,\ldots,x_n)$ into the Fmid-domain is simple and straightforward using the \texttt{rank} command in  R. 

\beq \label{R:1}
u \leftarrow  \dfrac{\left[\texttt{rank}(\mathbf{\xb},\texttt{ties.method = c("average")}) -.5\right]}{n} .
\eeq

%%%%%%%%%%%%%%%%%%%%%%%%%%%%%%%%%%%%%%%%%%%%%%%

\begin{table}
 \caption{Mid-Rank transformation of Protime variable (First 10 samples)}
\label{table:fm}
\centering
\begin{tabular}{c||ccccccccccc}
$\xb$ (Raw Data) &80 &75 &85 &54 &52 &78 &46 &63 &85 &62 &$\ldots$ \\
\hline
$\ub$ (Mid-Rank) & 0.778 &0.732 &0.818 &0.392 &0.357 &0.767 &0.261 &0.534 &0.818 &0.511 &$\ldots$ \\
\end{tabular}
\end{table}

From now on we will work with the $U$ matrix rather than the $X$ data matrix.
\bea \label{eq: XU}
\big[ \texttt{Original Raw Data}\big] &\rightarrow& \big[ \texttt{Mid-Rank Transform Data}\big] \nonumber \\
X &\rightarrow& U 
\eea
\subsection{Two Sample Data Analysis \& Comparison Density} \label{cd}
\noi \textit{Notation}. 

 To motivate the concept of ``Comparison Density'', we consider the two population classification problem. We assume that  $(X_i,Y_i)$, $i=1,2, \ldots,n$ are independent observations where $Y$ is a binary $0-1$ variable corresponding to the two populations from which we are observing the $p$ dimensional random vector $\X$, with corresponding sample sizes $n_0$ and $n_1$ (say). 

We use the notation $F(x|Y=y)=\Pr(X \le x|Y=y)$ to denote the conditional distribution of a single variable $X$, a generic component of the vector $X$ (e.g., $X=$ Bilirubin level in Panel A of Figure \ref{pic:intro}  ) given $Y=y$, where $y \in \{0,1\}$. For notational simplicity, from now on, set $F(x|Y=1)=F(x)$ (distribution of ``red'' class) and $F(x|Y=0)=G(x),x \in \cR$. Let $H(x)$ denote the unconditional cumulative distribution (cdf)
function which is given by
\[
H(x) = \pi F(x) + (1- \pi) G(x),  % pi = proportion of sample from first class
\]
where $\pi=\Pr(Y=1)$. For simplicity of exposition, suppose that $F(\cdot)$ and $G(\cdot)$ are absolutely continuous with quantile functions  $F^{-1}(\cdot)$ and $G^{-1}(\cdot)$, respectively.

\noi \textit{Comparison Distribution Function}.

Under the classification setup for variable selection, our main objective is to compare the two distributions $F$ and $G$ and to check whether they differ \textit{significantly}. Based on that, we declare whether the variable $X$ is informative or not. To compare two probabilities $p_1$ and $p_2 \in (0,1)$ we prefer to use $p_1/p_2$ rather than $p_1-p_2$. Applying this philosophy for the  purpose of comparing two distribution functions $F(x)$ and $G(x)$, we propose the \textit{ratio comparison principle} and define concepts of \textit{pooled} \textit{Comparison Distribution function}  \citep{parzen83a,parzen92} as 
\beq
D(u;H,F) = F(H^{-1}(u)), ~ 0 < u < 1.
\eeq
Its density, called the \textit{Comparison Density function}, satisfies
\beq
d(u; H, F):= D'(u;H,F) = \dfrac{ f(H^{-1}(u)) }{ \pi f(H^{-1}(u)) + (1-\pi) g(H^{-1}(u))}, ~ 0 < u < 1.
\eeq
Comparison distributions and density concepts can be used to compare two discrete distributions ( with pmfs $p_{F}$ and $p_{H}$) of a variable (e.g., for the variable \texttt{age} in hepatitis data) by setting
\beq
d(u; H, F) = \dfrac{ p_{F}(H^{-1}(u)) }{ p_{H}(H^{-1}(u))},~ D(u;H,F) = \int_0^{u} d(s,H, F) \dd s,  ~ 0 < u < 1.
\eeq

\noi \textit{PP-Plots.}

Suppose that $H$ is discrete and $x_1 < x_2 < \ldots < x_r$ are the probability mass points of $H$. Then one can easily verify that $D(u;H,F)$ is a piecewise linear between its values at $u_j=H(x_j)$ and that
\[ D(u_j;H,F)= F(H^{-1}(u_j)) = F(x_j), ~\text{for } j=1,2,\ldots,r. \]

The graph of $D(u;H,F)$ is called the PP-Plot. It joins the points $(H(u_j),F(u_j)), j=1,2,\ldots,r$ by linear interpolation.\\
             
\noi \textit{Why is PP-Plot Important for Variable Selection ?}

First note that identifying an important variable $X$ and testing $H_0: F(x)=H(x)$ are equivalent problems. If $F \approx H$, it indicates that we cannot hope to extract any meaningful information from this particular variable. Now, to see that it is  also related to the comparison distribution make the change of variable $u=H(x)\text{ or } x=H^{-1}(u)$, to express the hypothesis to be tested as
\[ H_0:F(H^{-1}(u)) = u, \, \,\text{ or } \, H_0:D(u_j;H,F)=u. \]
 Figure 3. shows how this hypothesis can be visually tested using the PP-Plot. Looking at the PP-plot of Panel B ( Figure 3), we can conclude that gene \# 610 of the prostate cancer data is probably an uninteresting variable. We can therefore measure \textit{importance} of a variable in terms of the norm of $D(u) - u $ or $d(u) - 1$ or by other functional of the comparison density. To make this concept more precise, we introduce a class of specially designed score functions and an orthogonal series model for $d(u)$.

\subsection{Estimating the Comparison Density}
% \begin{proportion}
%  The comparison density integrates to 1:  $\int_0^1 d(u) du =1$
% \end{proportion}
In this section we construct an estimator of comparison density using \textit{orthogonal mid-distribution score functions} as basis functions, which will be used heavily to build measures of importance and thus  for identifying interesting variables. In the context of variable selection, the sole purpose of the comparison density is to indicate the \textit{nature of discriminatory information} hidden in a variable. The various shapes of the comparison density gives answers to the questions \textit{why} \& \textit{how} a variable is important and indicate direction for follow-up scientific investigation. For example, a quadratic pattern (Panel A of Figure 4(b)) of the comparison density would suggest that the variable has high second order information, which means it shows highly significant difference in variability between two classes. It is the job of the biologist to explain this statistically significant discriminative pattern (see Remark 1).

%---------------------------
\begin{figure*}[tpbh] 
\vspace*{-.1in}
\begin{center}
\includegraphics[width=8.1in]{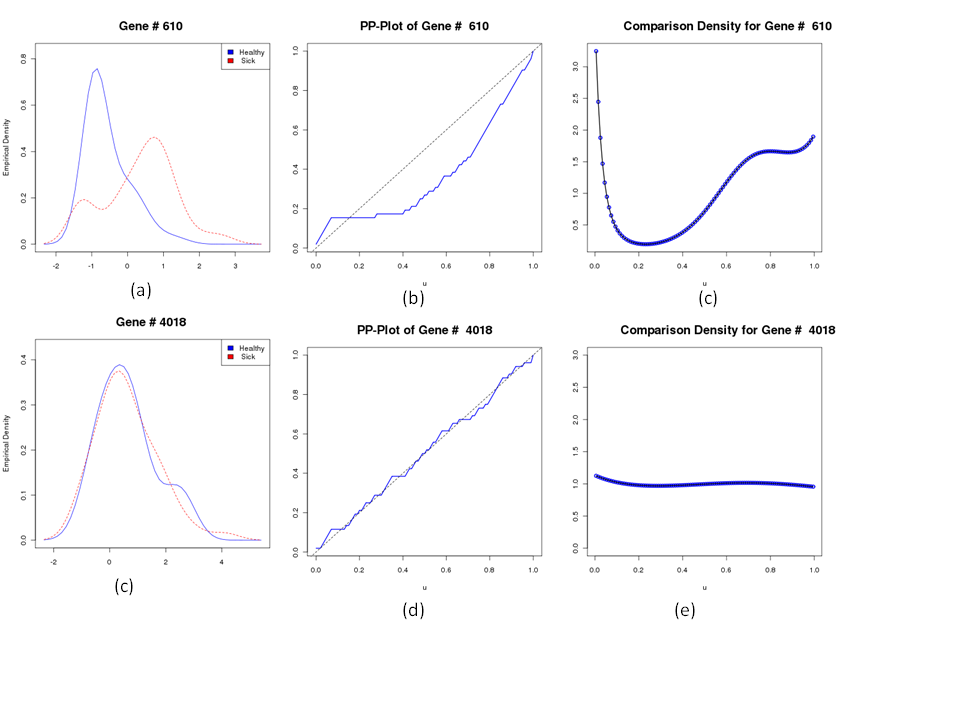}\end{center}
\vspace*{-1.17in}
\caption{\emph{PP-Plot and Comparison Density : Plot (a) and (b) denotes the two class distribution of Gene \#610 and \# 4018 of the Prostate cancer data set. Corresponding PP-Plots are shown in (b),(d) which is the plot of $\wtD (u; H,F) = D(u;\wtH,\wtF )$. Plot (c) and (e) the the corresponding smooth comparison density estimate $\dhat(u)$. The flat comparison density of plot (e) indicates that the distribution of Gene \# 4018 barely vary over the two classes which leads to the conclusion that Gene \# 4018 contains no information for class and can be considered as noise. Whereas Gene \# 610 shows a considerable departure from uniformity and its shape indicates that it carries second and higher order information.} } \label{pic:cd}
\end{figure*}

\noi \textit{Orthogonal Series Density Estimation of the Comparison Density}
  
We consider the following expansion of comparison density  
\beq
d(u) := \sum_{k=0}^{\infty} \te_k S_k(u), \: 0<u<1,
\eeq
where $\{S_k(u) \}_{k=1}^{\infty}$ forms a complete orthonormal basis for $L^2[0,1]$. So that
\[\te_k = \int_0^1 S_k(u) d(u) \dd u = E(S_k(U)),\]
since $\int_0^1 d(u) \dd u =1$. In practice, the comparison density can be approximated by a finite series

\beq
\dhat(u;\te) =  \sum_{k=0}^{M} \hte_k S_k(u), \: 0<u<1, 
\eeq
for a suitable choice of $M$, called truncation point.

% \beq 
% d(u;\teb)= \exp\big ( \sum_{k=1}^M  \theta_k \phi_k(u) - \Psi_M(\teb)\big),\quad 0<u<1
% \eeq
% where $\{\phi_k\}$ are orthogonal basis functions  (or score functions) for $L^2[0,1]$. This model was First introduced by Neyman (1937) in the context of goodness of fit test. 
\noi \textit{Motivation: How to design score functions}

A representation of the score coefficient is given by,
\bea \label{eq:theta}
\te_k &=& \int_0^1 d(u) \, S_k(u) \, \dd u \, =\, \int_0^1 S_k(u)\, f(H^{-1}(u))/h(H^{-1}(u)) \, \dd u  \nonumber \\
&=& \int_{- \infty}^{\infty} S_k(H(x)) f(x) \, \dd x \,= \, \Ex \left [ \, S_k(\,H(X)\,) \:|\: Y=1 \right],
\eea 
%Where $R_1,R_2,\ldots R_{n_1}$ denote the ranks of class one under pooled sample. So, we can estimate the score coefficients by $\te_k = 1/n_1 \sum_{j=1}^{n_1} S_k(R_j/n).$

The two sample Wilcoxon rank-sum test $1/n_1 \sum_{j=1}^{n_1} R_j$ is expressed in terms of the mid-rank transformation ($\Hmid(\cdot)$) or $(R_j - .5)/n$, with mean $.5$ and variance given by (see \cite{parzen04b} for elegant proof)
\beq \label{eq:var}
\var \left[ \Hmid(X_k) \right] \,= \sigma_{mid;X_k}^2 = \, \frac{1}{12}\left(1-\sum_a | p(x'_a;X)|^3 \right),
\eeq
where $x'_i$ denotes the $i$th distinct value of $X$. We use the following representation of Wilcoxon statistics in terms of mid-rank transformation, define
\beq \label{eq:wil1}  \Wil = n_1 \sum_{j=1}^{n_1} \left[ \, (R_j - .5)/n \,  \right] \,=\, \: \tilde \Ex \left[\, \tHm (X) |Y=1  \, \right] \eeq
Eq. \eqref{eq:var} and \eqref{eq:wil1} motivate us to define score function $S_1(u) = \sigma_{mid}^{-1} \, (u - .5), \; \text{where } u = \Hmid(x)$, to represent a version of the Wilcoxon statistic (linearly equivalent version or normalized, which is asymptotically normal as a score statistic, given by $\int_0^1 S_1(u) \, \wtd(u) \dd u$, where $\wtd(u) = d(u; \wtH, \wtF)$, a raw estimator of $d(u)$. This new interpretation of Wilcoxon statistics in terms of a (linear) f\textit{unctional of the comparison density} with appropriately chosen score function, opens up new possibilities to design effective score functions to build powerful variable selection detectors, which we will explore in details in the following sections.

\noi \textit{Construction of score functions.} 

Novelty of our approach is in the construction  of the  basis functions. In contrast to the standard practice of taking the basis as powers of $x$, here we construct  orthonormal  score functions based on  ranks through mid-distribution transform. Define
\beq \label{eq:score}
 S_1(x) = \frac{\left( \Fm(x) - 1/2\right) }{\sigma_{mid}} , ~ x \in \cR 
  %\ts_1(u) &=& s_1(\tilde \Hp^{-1}(u)) , ~~ u_{k-1} < u < u_{k} \label{mid-score}
\eeq
and then sequentially construct $S_k(x) , \kM $ by Gram-Schmidt ortho-normalization
 of $S_1^k(x),\kM $. 
% In the next section we will describe how the score functions will be used for variable selection.

% \noi \textit{How to Compute Score Functions.}
% 
% Generating score functions for a any variable X is very simple and straightforward. We need a one line R-code in conjunction with \eqref{R:1}
% 
% \beq \label{R:2}
% S ~\leftarrow~ \texttt{poly}(S,\texttt{df}=M) ,
% \eeq
% where $S=(u-\texttt{mean}(u))/\texttt{sd}(u)$.

%-------------------
\begin{figure*}[tpbh] 
\vspace*{-1.4in}
\begin{center}
\includegraphics[width=8.2in]{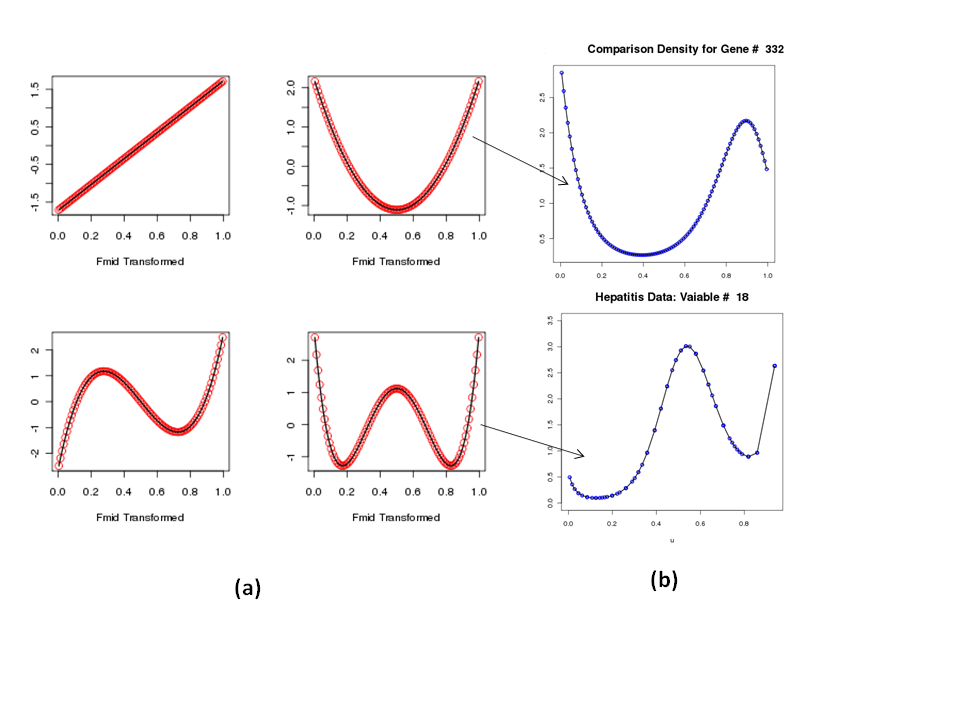}\end{center}
\vspace*{-1.17in}
\caption{\emph{Shapes of Score functions and the induced Comparison density  : (a) describes the shapes of the first four score functions $S_1(u), \ldots, S_4(u).$ (b) $L_2$ comparison density estimate using the four score functions. The shape of the $\dhat(u)$ for Variable \# 18 indicates that it has a information in the \textit{tail behavior} (or the fourth order moment)}.} \label{pic:score}
\vspace*{-.6in}
\end{figure*}

\section{Interpretation and Insight}
The usefulness of $d(u;H,F)$ comes from the following fact

\beq \label{eq:flat}
d(u;H,F) =1 ~\text{ iff }~ F(x)=G(x), \text{ for all } x.
\eeq
Figure 3e illustrate this behavior.
% Also it is a easy exercise to show that $$ \rm{KL}(F \pl G) = 0~ \text{ iff }~ d(u;H,F)=1.$$ 
% 
% 
%  It is somewhat similar to the fact of the spectral density of white noise. 
The \textit{flat} comparison density indicate that the variable is pure noise as a detector. But the more interesting phenomena is the case where the variable is influential. In that case $d(u)$ takes varieties of interesting shapes.
% Equation \eqref{eq:flat} imply that the  $\| d(u)-1  \|^2$ measure how much a variable is informative about the class. This is further confirmed by the following theorem.
The following theorem gives a justification of our score statistics which will be defined soon,
\bthm[\textit{Goodness of Fit Interpretation}] The variable importance can be quantified by,
 \beq 
\| d(u)-1  \|^2 = \int_0^1 d^2(u) \dd u -1 = \sum_{k=1}^{\infty}\theta_k^2 \eeq
\ethm \label{eq:gof}
\bpf
Proof directly follows from Parseval's identity.
\epf
Here $\theta_k$'s are the $L_2$ parameters of the density. It states that $\sum_{k=1}^M \hte_k^2$, with a proper choice of $M$, acts as a measure of how uniform the comparison density is, which in turn assigns an importance measure to a particular variable.
 The clever construction of our basis function opens another important interpretation in terms of nonparametric rank correlation. $\hte_1$ can be neatly rewritten as the following which add a nonparametric flavor to it. 
% Now the following result (i) we don't need to estimate the $d(u)$ to get (ii) The nonparametric flavor was attached due to the design of the score function according to our recipe \eqref{eq:score}.

% %%%%%%%%%%%%%%%%%%%%%%%%%%%%%%%%%%%%%%%%%%%%%%%
% 
% 
% \begin{figure*}[tpbh] 
%  \centering
%  \vspace*{-.5in}
% \subfigure[Mean Difference]{\includegraphics[width=3.2in]{mean.png}}
% \subfigure[Scale Difference]{\includegraphics[width=3.2in]{scale.png}}\\
% \subfigure[Contrasting Skewness]{\includegraphics[width=3.2in]{skew.png}}
% \caption{\emph{ Ranked under different category :}(a) ; (b) ; (c) .} \label{pic:3order}
%  \vspace*{-.1in}
%  \end{figure*}
% 
% %%%%%%%%%%%%%%%%%%%%%%%%%%%%%%%%%%%%%%%%%%%%%%%%%%%%%%%%%%%%%%

\bthm[\textit{Rank Correlation Interpretation}] Let $\corr [Y,S_1(\,\Hmid(X)) ] = \R [Y,S_1(\,\Hmid(X)) ]  $. Then we have the following correlation interpretation
\beq \label{eq:wil}
 \hte_1 = \sqrt{(1-\pi)/\pi}~\R\left [Y,S_1(\,\Hmid(X) ) \,\right] = \Wil (X) \eeq
\ethm
\bpf
Note that, 
\bea 
\R\left [\,Y,S_1(\,\Hmid(X) ) \,\right ] &=& \Ex\left [ \,S_1(\,\Hmid(X) )\,Y  \,\right ]/\sqrt{\pi \, (1-\pi)} \nonumber \\ 
&=& \sqrt{\dfrac{\pi}{1-\pi}}\: \Ex\left [ \,S_1(\,\Hmid(X) )\, | \, Y=1  \,\right ] 
\eea
Comparing this with Eq. (2.9) completes the proof.
\epf

Implications of this result are :  (i) The Wilcoxon statistic can be interpreted as a correlation between Y and $S_1(X)$. In this sense, it is a linear detector. (ii) The estimate of $\te_1$ has a nonparametric rank correlation interpretation by virtue of the special choice of the basis functions that we have introduced in the previous section.

The representation \eqref{eq:wil} motivates us to propose  Criterion Correlation(CR) statistics, which is an important generalization, yielding a general \textit{nonlinear} rank based detector:
\beq \label{eq:P}
CR = \sum_{a=1}^M \Big|\, R \left (Y,\,S_a(\Hmid(X))  \,\right ) \,\Big|^2.
\eeq

\subsection{Component Correlation Interpretation}
The measure that we have introduced in Eq \ref{eq:P} have a third interpretation from component correlation perspective. To define what we define by components, let's first choose score functions $S_1(u), S_2(u),\ldots S_M(u)$ satisfying 
\[ \int_0^1\,S(u)\,\dd(u)=0, ~\text{ and }~ \int_0^1\,S^2(u)\,\dd(u)=1. \] 
Now we form the components or linear detectors $j=1,\ldots,M$ by 
\beq
\wtT(S_j) = \wtE \left( S_j(u)\right) = \int_0^1\,S_j(u)\,\dd (\wtD(u)-u) = <S_j,\wtd>,
\eeq
which is often shown to follow asymptotically normal, under $H_0$. It is easy to verify using Eq (2.11) and Theorem \ref{eq:wil} that the Wilcoxon statistic (linearly equivalent)
has a following equivalent representation in terms of \textit{functional} of the comparison density empirical process,
\beq
<S_1(u),d> ~ = ~\int_0^1\,S_1(u)\dd (\wtD(u)-u),
\eeq
where $S_1(u)=\sigma_{mid}^{-1}(u-.5).$  Under this new representation, Wilcoxon rank-sum test can be viewed as just the \textit{first component} of our proposed test statistic (Eq \ref{eq:P}). Further more, many important classical test statistics to test the equality of distributions ($H_0:F(x)=H(x), \forall x $) can be represented as weighted sums of squares of suitable components (which is also known as a quadratic detector):
\[ \sum_{j=1}^{\infty} \left\{ w_j<S_j,d> \right\}^2 .\]
For example $w_j=1/(j \pi)$ and $S_j(u)=\sqrt{2}\cos(j\pi u)$ gives the famous Cramer-Von Mises statistic. One reason to prefer this form of expressing nonparametric statistics is that they help to identify \textit{sources} of significance. The components can tell us \textit{how} the behavior of a specific variable is different in two different classes. The key concepts is the comparison distribution and score function which enable us to unify and choose between diverse statistics available for variable selection. Apart from integrating the different concepts, we are now in a position to deliver valuable insight about the operation of significant variables which could have significant impact on scientific understanding. In contrast conventional off-the-shelf variable selection machinery like t-statistics, Lasso, Wilcoxon statistics, simple Pearson correlation measure, etc. have a limited practical utility in the line of our intended application. In the next section, we will discuss few properties of the key stochastic process, comparison distribution empirical process (introduced in \cite{parzen83a}) to derive asymptotic distribution of our proposed variable selection statistic.

\section{Comparison Distribution Empirical Process: Weak Convergence and Limit Theory} \label{CDEP}
Rigorous starting point for our main result is provided by the following fundamental theorem for two sample comparison density empirical process \citep{ps68,parzen99}.
\bthm[\textit{Comparison Density Empirical Process}] \label{thm:CD}
Assume that $F$ and $G$ respectively have positive continuous derivatives $f$ and $g$ respectively and that also $\lim_{\nti}(n_1/n) = \pi \in (0,\infty)$. Suppose further that $d(u;H,F)$ and $d(u;H,G)$ are bounded on any $(a,b) \subset (0,1)$. We denote by $\cB_{F}$ and $\cB_{G}$ two independent uniform Brownian Bridges, given by $\cB_{F}(u;H,F) = \cB_F( \,F(H^{-1}(u))\,)$ and $\cB_{G}(u;H,G) = \cB_G( \,G(H^{-1}(u))\,)$. Then , as $\nti$,
\beas 
&&\sqrt{n}\left(D(u;\wtH,\wtF)  - D(u;H,F) \right)~ \rarrow^{d}  \\ 
&&(1-\pi) \left\{ \dfrac{1}{\sqrt{\pi}} d(u;H,G)  \cB_{F}(u;H,F) \,-\,\dfrac{1}{\sqrt{1-\pi}} d(u;H,F)  \cB_{G}(u;H,G)  \right\}, \, 0<u<1. \eeas
\ethm
Note that under $H_0: F=G=H$, $\sqrt{n}\left(D(u;\wtH,\wtF)  - D(u;H,F) \right)$ turns out to be $\sqrt{n}(\wtD(u)-u)$, where $\wtD(u)=D(u;\wtH,\wtF)$. Our aim here is to derive the asymptotic distribution of the $\int_0^1\,J(u)\dd (\wtD(u)-u)$, for suitable linear functional of comparison density empirical process. 
\bthm[Asymptotic Distribution] \label{thm:asymp}
Define $\La(J) = \sqrt{n} \left \{ \int_0^1 J(u)\dd \wtD(u) -  \int_0^1 J(u)\dd D(u)  \right\}$. Then under $H_0$, $\La(J)$  is asymptotically normal with covariance kernel $K_{\La}(J_1,J_2)$, is given by
\[K_{\La}(J_1,J_2)~ = ^{H_0} ~\left(\dfrac{1-\pi}{\pi}\right) \left\{ \int_0^1\,J_1(u) J_2(u) \dd u -  \int_0^1\,J_1(u) \dd u  \int_0^1\,J_2(u) \dd u \right\}\]
\ethm
For proof see \cite{parzen83a}.
This results readily gives us the following useful corollary for the limit distribution of our rank based statistics.
\begin{coro}[\textit{Null Density}] \label{thm:nd}
 $n \,P \, \rarrow^d \, \chi_M^2. $
\end{coro}
\bpf
Our designed score functions are orthonormal, i.e., in our case $\int_0^1\,J_1(u) J_2(u)\dd u =0$ and $\int_0^1\,J_1^2(u) \dd u = 1$. This implies that $\R\left [\,Y,S_1(\,\Hmid(X) ) \,\right ]$ are asymptotically iid and
\beq \sqrt{n} \, \R\left [\,Y,S_1(\,\Hmid(X) ) \,\right ] \, \rarrow^d \,  \cN(0,1). \eeq
Result follows from the definition of the CR-statistics (Eq. \ref{eq:P} ).
\epf
\section{Algorithm and Illustration } 
Before going any further, at this point we will take a pause and revisit the Prostate data example to see how to apply the previous theory to come up with an importance score using the CR-statistic for each of the $6033$ variables.
\begin{description}
 \item[(i).]\textit{Transformation.} The first step is to transform raw data matrix $X$ to $U$, mid-rank transformed values ( see Eq 2.3). 
\item[(ii).] \textit{Basis Expansion.}  Construct score functions $S_k(\,\Hmid(X_k) \, ), \: k=1,\ldots,4$ for each variable. Panel A of Figure 4 shows the generic shapes of first four score functions for the prostate cancer data. This score functions can be thought of as a kernel for projecting the raw data from $p$ dimension to $4\,p$ dimension. 
\item[(iii).] \textit{Computing CR-statistic.} Once we compute the sufficient statistics (score functions) for each features, we use Eq. \ref{eq:P} and generate the CR-statistics for each variable.
\item[(iv).] \textit{Ranking \& Categorizing.} We can rank the variables according to the values of the CR-statistic (see Fig 2) and select a \textit{`proper threshold'} to identify   few influential variables from the list of $6033$ variables. More importantly we can categorize the important variables according to  their \textit{`discriminatory role'} played by the variables. Table  \ref{table:pro} not only finds the top interesting variables but also label the variables according to `what variable is contributing to what type of information'.
\item[(v).]  \textit{Interpretation.}  Lastly plot the comparison density estimate for the top few interesting variables to demonstrate graphically `how and why' those variables are important. The various shapes of comparison density (see Figure 3 (c,d) and 4 (b)) conveys `why' a particular variable is interesting for scientific understanding.
 \end{description}

What's remain is the following question: `` How many top variables to select ? ''. The next Section is dedicated primarily to build a simple yet powerful theoretical set up to attack this problem.

% \subsection{Hepatitis Data}
% 
% \begin{center}
% \textbf{WILL BE ADDED}\end{center}
% %%%%%%%%%%%%%%%%%%%%%%%%%%%%%%%%%%%%%%%%%%%%%%%

\begin{table}
 \caption{Top ranked variables under different Category: Prostate Cancer Data}
\label{table:pro}
\centering
\begin{tabular}{c|c|c|c|c|}
$R_1^2$(Mean) & $R_2^2$ (Variance) & $R_3^2$ (Skewness)& $\sum_{i=1}^2R_i^2$ & $\sum_{i=1}^3R_i^2$ \\
\hline
\hline
452 &614 &16 &614 &332 \\
411&1546 &669 &77 &77 \\
739&377 &423 &{\bf 332 } &614 \\
4552& 1139& 24& 808&579 \\
\hline
\end{tabular}
\end{table}

% \begin{table}
%  \caption{Top ranked variables under different Category: Hepatitis Data}
% \label{table:H}
% \centering
% \begin{tabular}{c|c|c|c|c|}
% $R_1^2$(Mean) & $R_2^2$ (Variance) & $R_3^2$ (Skewness)& $\sum_{i=1}^2R_i^2$ & $\sum_{i=1}^3R_i^2$ \\
% \hline
% \hline
% 17 &14 &17 &17 &17 \\
%  12&17 &16 &{\bf 18} &{\bf 18}\\
% 18& 18& 15& 14&14 \\
% \hline
% \end{tabular}
% \end{table}

%======================================================================================================================================
%===================== SECOND PART   =====================================================================================================

\section{Separating Signal from Noise} \label{sec:FDR}

\noi \textit{Problem Statement} 

 The problem that we want to tackle now, has a very general setup, which is commonly known as detection/threshold selection problem.
 Figure \ref{pic:thres} shows the plot of the sorted values of the CR-statistic and the problem is to select a proper threshold. To control the number of falsely rejected hypotheses,
\cite{BH95} introduced the false discovery rate (FDR) and described a procedure
to control FDR. There are many variants of FDR that have been introduced in the past decade (mFDR,pFDR,fdr). \cite{efron04} introduced an alternative measure local false discovery rate (Locfdr/fdr) based on a two-group model to control FDR from a density estimation perspective. 
The local false discovery rate is defined as 
\beq \label{eq:fdr}
\fdr(z) ~\equiv ~\Pr\{ \rm{null} \mid Z=z\}~ =~ p_0 \dfrac{f_0(z)}{f(z)},
\eeq
where $\Pr{\rm{Null}} = p_0$ , $f_0$ is the theoretical null density, $f_1$ density of interesting variables and $f(z) = p_0 f_0 + (1-p_0)f_1$, pooled density.
Recently, \cite{Om10} proposed Mixfdr method, based on the hierarchical empirical Bayes mixture model to accurately estimate fdr. A few key observations on the fdr literature are: 

(i) The main focus of current research on fdr concerns flexible estimation of $f(x)$, using either exponential model or splines or fancy mixture models. 

(ii) At the second stage, the estimator for $\fdr(z)$ is constructed (Eq \ref{eq:fdr}) taking the ratio $f_0(z)/\hf(x)$, where $f_0$ is the theoretical null and assuming $p_0=\Pr(\rm{Null})=1$, without much harm (cf. \cite{efron01,efron04} for details). Broadly speaking, the estimation of $\fdr$ is accomplished in a two steps, first estimating the $f(x)$ and then taking the ratio. 

As we shall argue in the next section that there is at least \textit{two fundamental flaws} with this \textit{two-step} approach for $\fdr$ estimation. First, we look at some simulation example.
 
% Clearly the problem of recovering signal from the noise is not always feasible. If the signal strength is weak , i.e, if the separation between signal and noise is not sufficient enough we cant hope to detect the signals. In that sense this is exactly the same problem as bump hunting. We need certain minimum separation between two bumps so that we can even detect that otherwise there is a possibility that we might smooth out the bumps and find no significant bumps/signal. Similarly this is exactly in parallel with the modeling mixture density. It was noticed that we need a minimum separation to detect the two mixture component successfully.

\begin{figure*}[tpbh] 
\vspace*{-.1in}
\begin{center}
\includegraphics[width=5.5in]{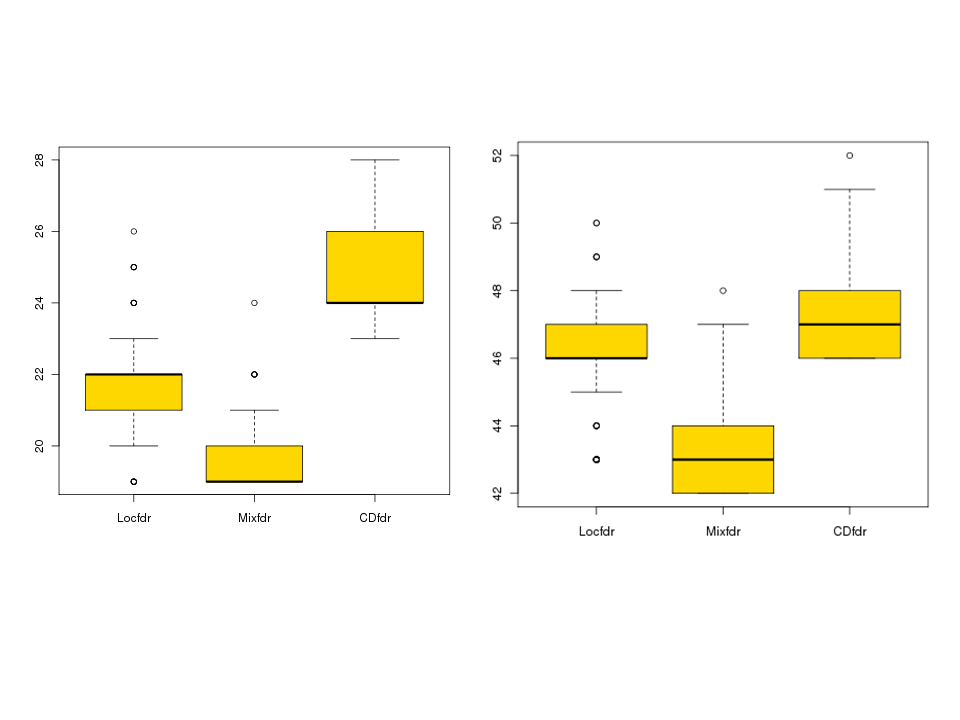}\end{center}
\vspace*{-.8in}
\caption{\emph{Gaussian Shift model : Left panel M=25, $z_i$ was drawn from $\cN(4.52,1)$ (once for all). Where as the Right panel has 50 $z_i$ generated from $\cN(4.52,1)$ (once for all). The figure shows the boxplot of the number of $\mu_i$ was selected as nonzero over $100$ simulation run.}} \label{pic:GS}
\vspace*{-.05in}
\end{figure*}

\begin{example}[\textit{Gaussian Model with Contamination}]
Consider $z_i \sim \cN(\mu_i,1), i=1,2,\ldots 1000$ and we have M nonzero components $\mu_i=4.52$. This example is similar to the one considered in \cite{donoho06}. We have generated M non-zero significant components once for all and then at each simulation iteration we have added additional $(1000-M)$ noise and the task is to recover the important signal for the corrupted version.
Figure \ref{pic:GS} illustrates the result of a simulation study for Gaussian shift model. It compares three competing methods for threshold selection (i) Locfdr \citep{efron04}, (ii) Mixfdr \citep{Om10} and comparison density based fdr(CDfdr). It is evident that CDfdr does a better job of finding the true number of nonzero components compared to other two methods and adapts quite successfully to the underlying sparsity level. A surprising fact is that the two competing methods consistently underestimate the true number. A possible explanation of this false negative phenomena and a simple solution to it is explained in Section 6.1. 
\end{example}

%-------------------

\begin{example}[\textit{Non-Gaussian Model with Contamination }]
Consider the problem where the $M$ nonzero signals were drawn from $\text{Uniform}[2,4]$ (once and for all) and $(1000-M)$ Gaussian white noise were added to it.
At each simulation run, we generate random noise and we mix it with the signal. This example was used in \cite{Om10}. Our aim to study how successful these three methods are for finding out the true number of significant signals and as a next step, investigate few plausible reasons for under-performance.  Figure \ref{pic:score} demonstrate the out-performance of our method in contrast with other two methods.

The comparison density based fdr (CDfdr), which we will describe in the following section appears to be the most reliable and straightforward technology for the variable selection in the examples above. The objective of the next section is to explain how we derive and compute the CDfdr. In fact, the nonparametric approach developed here, seems applicable to a wider class of detection problems having non-gaussian noise. 
\end{example}

%-------------------
\begin{figure*}[tpbh] 
\vspace*{-.2in}
\begin{center}
\includegraphics[width=4in]{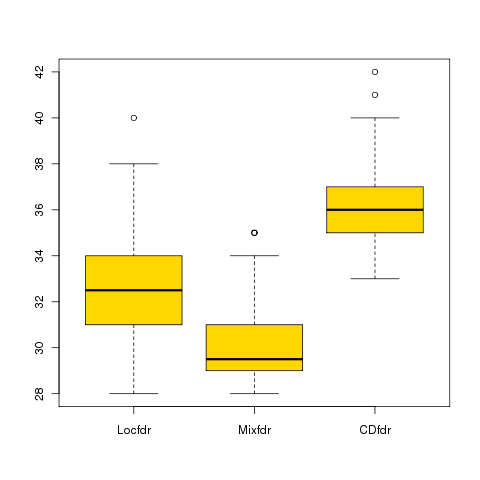}\end{center}
\vspace*{-.5in}
\caption{\emph{Non-gaussian Model with Contamination: Three boxplots comparing the performance of variable selection efficiency over $100$ simulation run} .} \label{pic:score}
\end{figure*}

\subsection{Rationale \& Main Contributions} 
The first crucial observation is that estimating \textit{directly} the ratio of two density is much more efficient and straightforward rather than estimating them separately and then taking the ratio. (i)\textit{One sample Comparison density:} As a remedy, we apply idea of comparison density introduced in Section 3 in one sample case, as a means of devising a direct procedure to estimate the \textit{proper} ratio of the two density. We prefer to work with the inverse-fdr, $f/f_0$ as opposed to $f_0/f$. There is another factor that merits our attention.
(ii)\textit{Pre-flattened smoothing} (\cite{parzen79}) Even if we reduce the estimation ratio of two density into a single step process, the traditional exponential density estimation typically does not work. This is due to the fact that $f$, the pooled density is fat tailed compare to the null distribution $f_0$ which makes \mbox{Support($f_0$)} $\subset$ \mbox{Support(f)} and introduce extra challenge. This directly affects the tail portion of $f/f_0$. 
Efficient estimation of the tail probability depends upon reducing the dynamic range (sharp peak) at the corners and this calls for some new techniques to accurately estimate the tail portion. To get an estimate of the fdr, we are only going to focus on the tail estimation, rather than estimating $f/f_0$ on the full support domain, as the central part will not contribute anything to estimate the fdr for significant variables. 

To our knowledge, this connection between tail of Comparison density and fdr estimation is new and opens many avenue for further investigations. 

%-------------------
\begin{figure*}[tpbh] 
\vspace*{-.2in}
\begin{center}
\includegraphics[width=7.2in]{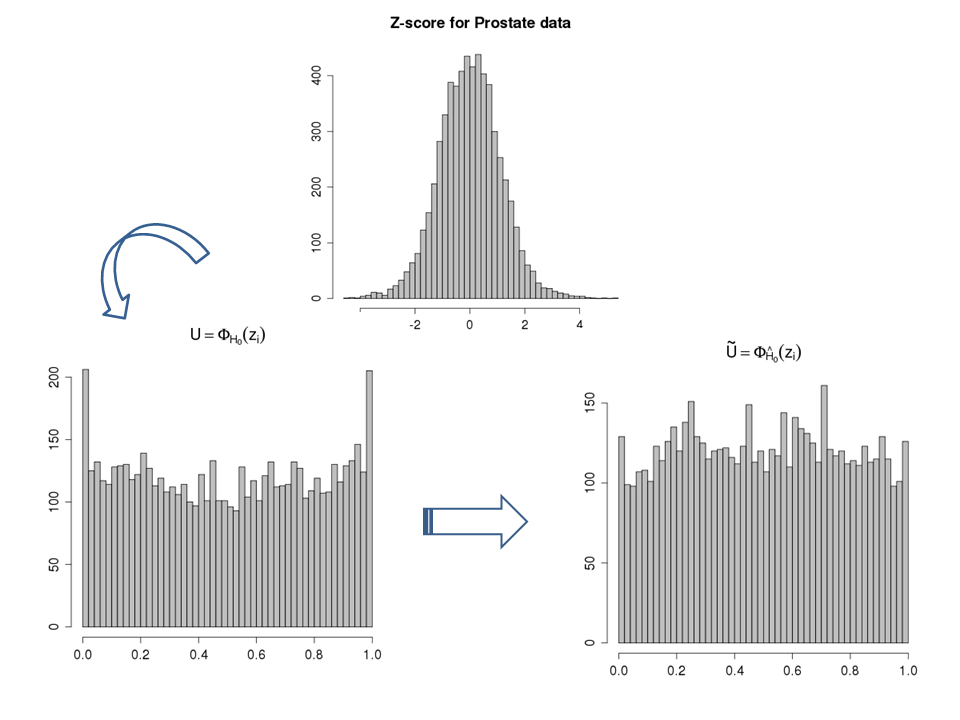}\end{center}
\vspace*{-.3in}
\caption{\emph{CDfdr Algorithm :} \textit{We transform the raw Z-score vales to $U_i=\Phi_{H_0}(z_i)$. The distribution of $U_1,U_2,\ldots U_p$ directly estimate the ratio $f(\Phi_{H_0}^{-1}(u))/f_0(\Phi_{H_0}^{-1}(u))\, = \, 1/\fdr(u)\, = \, d(u)$. Directly estimating the density of $\Phi_{H_0}(z_i)$ is difficult due the the presence of the sudden sharp peak at the both end (which signifies the presence of the signal). So instead we perform pre-flattening transformation using $ \widetilde U_i=\Phi_{\hat H_0}(z_i) = \Phi \left[ (x- \hat \mu_0) \, /\hat \sigma_0) \, \right]$. Where the parameters are estimated from the pooled data. After this transformation we have a better chance to model it efficiently as it reduces the dynamic range at the boundary considerably.  } } \label{pic:GOF}
\vspace*{-.2in}
\end{figure*}

\subsection{Comparison Density based fdr : Goodness of fit Problem} \label{sec:CDfdr}
\noi \textit{One sample Goodness of fit formulation}

Let $T_1,T_2,\ldots , T_p$ be random sample from $F$.
In Section 2.2 we have introduced two sample comparison density analysis and now we are going to introduce the one sample version of it. The basic task of estimating $\fdr$ starts with the hypothesis testing question that $F(t) = F_0(t)$, where $F_0$ is the specific distribution either known theoretical null or the empirically estimated version of it. The probability integral transformation $U=F_0(X)$ transforms the data $T_1,T_2,\ldots , T_p$, to a pseudo-data $U_1,U_2, \ldots, U_p$ on unit interval, which allows us to reformulate the problem into a problem of \textit{testing uniformity in a unit interval}. This is one of the most important canonical problem of statistics and is of independent interest.

\noi \textit{Example}

Bottom left panel of Figure \ref{pic:GOF} shows the histogram of $\Phi(z_i), i=1,2,\ldots,p$, where $z_i$ is the z-score for the $i$th gene in the Prostate dataset. There is a clear departure of uniformity specially in the two corners.

\noi \textit{Direct Approach to Ratio Estimation}

Here we introduce a novel method that automatically estimate the ratio of two density in a single step and render some useful modeling and analysis convenience. The key is the following simple and yet deep fact about comparison density $d(u)$ for one sample,
\begin{prop} \label{prop:GOF}
Density of $U$ is $f(F_0^{-1}(u))/f_0(F_0^{-1}(u))=d(u)$, where $u=F_0(t)$.
\end{prop}
\bpf
 To see why this is the case first look at the corresponding distribution function
\[\Prob(F_0(Z) \le u) = \Prob(Z \le F_0^{-1}(u)) = F (F_0^{-1}(u)),\]
which implies that $ \left( F(F_0^{-1}(u)) \right)' = f(F_0^{-1}(u))/f_0(F_0^{-1}(u)) = d(u).$
\epf

This is the main reason why we prefer to work with the inverse-$\fdr$. Now let's look at a simple algorithm to estimate the fdr based on the previous observation.

\noi \textit{Simple Algorithm}

\noi I. Transform $z \rarrow F_{H_0}(z)=u$.

\noi II. Estimate density of u. This will give us the ratio of two density as a function of u.

\noi III.  Find $\{u\,;\, \dhat(u) > 5\}$, as the conventional threshold for reporting signal/interesting variables is $\fdr(z) < .2$.

\noi \textit{Nonparametric Pre-whitening}\, \citep{parzen79}

Proposition \ref{prop:GOF} reduce the problem of fdr estimation into a single density estimation problem. But conventional nonparametric density estimation encounter roadblock. Bottom left panel of Figure \ref{pic:GOF} shows a typical shape of the p-values near the boundary. We expect this sudden peak in the variable selection scenario as the signal resides on the corner. Traditional nonparametric density estimators including exponential density performs poorly to capture the tail. Towards this, we may note that $\widehat{\fdr}(u)$ is specially important for $u$'s near the boundary as we expect to have significant variables at those regions. To overcome the challenge pose by this large dynamic range of the $\wtd$ near the corners we propose a nonparametric pre-flattened smoothing technique 
. Bottom right panel Figure \ref{pic:GOF} illustrate the stabilizing/flattening effect which is the histogram of $\widehat{F}_{0}(z)$. Here $\widehat{F}_{0}(z)$ can be interpreted as a pooled distribution under $H_0$. For prostate data, the mean and standard deviation for the pooled z-values turns out to be $0$ and $1.135$ and we choose $\Phi(x/1.135)$ as $\widehat{F}_{0}(z)$. We introduce artificially the $\widehat{F}_{0}(z)$ and write the ratio as, 
\beq
\dfrac{f(x)}{f_0(x)} ~=~ \dfrac{\widehat{f}_{0}(x)}{f_0(x)} \, \dfrac{f(x)}{\widehat{f}_{0}(x)},
\eeq

Note that the we have factorized the original ratio into two parts. In the first part there is no approximation error as we exactly know the function $f_0$ and $\widehat{f}_{0}$, so there is no estimation exercise. This works as a adjusting weight. Only the second part involves the density approximation. But now the good news is that, the simple trick of iterative density estimation through pre-flattening enables us to estimate the \textit{residual-ratio} $f(x)/\widehat{f}_{0}(x)$, much more efficiently compared to the original version $f(x)/f_0(x)$. In a broad sense our flattening technique could be considered as a \textit{regularization method} for estimating ratio of two density.

\begin{remark}[\texttt{Novel Empirical Bayes Connection}]
Note that our approach is free from any assumption of mixture model. It involves no tuning parameter and is a fully data analytic approach. As ratio of two density is a integral part of Bayes rule, it is conceivable that our novel rank based methodology for estimating the ratio in a fully nonparametric way opens a powerful implication and interpretation from empirical Bayes prospective.
\end{remark}

\subsection{Back to the Prostate Data}
In this section we will mention few findings for prostate cancer data and compare with other competing methods. fdr at the level .2 was used for threshold selection and $p_0$ was assumed to be 1. All of the method used the estimated empirical null as opposed to theoretical null model. For our case we have used the central matching estimation technique of \cite{efron04}. Table \ref{table:pros1} compares the variable selection performance of Locfdr, Mixfdr and CDfdr on the basis of Z-score. Whereas , Table \ref{table:pros2} compares the threshold selection procedure where we have use only the first two components of our CR-statistic (Eq. \ref{eq:P}). R packages ''locfdr`` and ''mixfdr`` was used to implement the Locfdr and Mixfdr methodology. 
%%%%%%%%%%%%%%%%%%%%%%%%%%%%%%%%%%%%%%%%%%%%%%%

\begin{table}
 \caption{Prostate Cancer data: Number of Variables selected using Z-score}
\label{table:pros1}
\centering
\begin{tabular}{c|c|c}
Locfdr & Mixfdr & CDfdr\\
\hline
\hline
 54& 49 & 46\\
\end{tabular}
\end{table}

%%%%%%%%%%%%%%%%%%%%%%%%%%%%%%%%%%%%%%%%%%%%%%%

\begin{table}
 \caption{Prostate Cancer data: Number of Variables selected using CR-statistic}
\label{table:pros2}
\centering
\begin{tabular}{c|c|c}
Locfdr & Mixfdr & CDfdr\\
\hline
\hline
 13 & 10 & 19\\
\end{tabular}
\end{table}

Using our similar methodology we can inspect the higher order variable screening and thresholding performance exactly in the same way.

\section{Discussion \& Future Direction}
Although the main goal of this article is variable selection the implicit aim is to demonstrate the effectiveness of quantile based comparison analysis as a means to generalize, unify different methodology for high dimensional inference. We made a effort to synthesize powerful applicable methodology combining several novel ideas of robust nonparametric statistics which can adapt for different data types. These development will possibly pave the way for modern analysis of high dimensional data using quantile, comparison density, mid distribution score function. One of the most attractive feature for our variable selection methodology is that, it generates explanation in terms of graphical tool. This graphical element of our method helps to translate \textit{numbers} to \textit{shapes} and thus act as a important diagnostic tool. 
In Section \ref{CDEP} we have studied the novel stochastic process, called two sample comparison density empirical process (introduced by \cite{parzen83a}) and studied the asymptotic properties of our statistic. It has shown that, linear rank statistics, the Cramer-von Mises, Anderson-Darling and many more can be studied in a unified way and can be conveniently represented as a functional of this fundamental process. In Section \ref{sec:CDfdr} we have neatly rewritten the fdr in terms of comparison density and established a novel connection with Empirical Bayes method. 
We have left untouched several important aspect including correlated variable selection, which is currently under investigation. However, we believe that some key ideas presented in this article might be extended much further than we have managed to do. For example, in trying to describe fdr we have introduced the Pre-flattened smoothing approach in conjunction with one sample comparison density concept. We can use this idea for a general purpose semi-parametric density estimation technique where we can choose any reasonable parametric distribution to be our flattering function. We hope the present article will contribute to a better understanding of variable selection from a broader perspective.

% \appendix
% %\section{Appendix}
% 
% \makeatletter   %% HAVE TO ADD SOMETHING HERE TO MAKE IT SAY "APPENDIX"
%  \renewcommand{\@seccntformat}[1]{APPENDIX~{\csname the#1\endcsname}.\hspace*{1em}}
%  \makeatother
% \section{Comparison Distribution Empirical Process}

%\bib

\end{document}